\documentclass{PoS}

\title{Status of the lambda lattice scale for the SU(3) Wilson gauge 
      action} 

\ShortTitle{SU(3) lambda lattice scale}

\author{\speaker{Bernd A. Berg}\\ 
        Department of Physics, Florida State University, Tallahassee,
        FL 32306, USA \\
        E-mail: \email{bberg@fsu.edu}}

\abstract{With the emergence of the Yang-Mills gradient flow technique
there is renewed interest in the issue of scale setting in lattice gauge 
theory. Here I compare for the SU(3) Wilson gauge action non-perturbative 
scale functions of Edwards, Heller and Klassen (EHK), Necco and Sommer 
(NS), both relying on Sommer's method using the quark potential, and 
the scale function derived by Bazavov, Berg and Velytsky (BBV) from 
a deconfining phase transition investigation by the Bielefeld group.  
It turns out that the scale functions are based on mutually inconsistent 
data, though the BBV scale function is consistent with the EHK data when 
their low $\beta$ ($\beta=5.6$) data point is removed. Besides, only the 
BBV scale function is consistent with three data points calculated from 
the gradient flow by L\"uscher. In the range for which data exist the 
discrepancies between the scale functions are only up to $\pm 2$\% of 
their values, but clearly visible within the statistical accuracy. }

\FullConference{The 32nd International Symposium on Lattice Field Theory,\\
		23-28 June, 2014\\
		Columbia University New York, NY}

\begin{document}

\section{Introduction} \label{sec_intro}

With the emergence of the Yang-Mills gradient flow technique \cite{L10}
there is renewed interest into the issue of scale setting in lattice
gauge theory. For a review see \cite{So13}. Therefore, it appears to 
be worthwhile to analyze the status of the lambda scale for the SU(3) 
Wilson gauge action from previous literature. Based on the Sommer scale
\cite{So94} there are two estimates (parametrizations) of the SU(3) 
scaling function, a paper by Edwards, Heller and Klassen \cite{EHK98}
(EHK) and another by Necco and Sommer \cite{NS02} (NS). Independently
an estimate of the SU(3) scaling function was later extracted by 
Bazavov, Berg and Velytsky \cite{BBV06} (BBV) using deconfining
transition coupling $\beta_t$ estimates and other information from 
a paper by the Bielefeld group \cite{Bo96} ($\beta=6/g^2$, where $g$ 
is the bare coupling of the SU(3) Wilson gauge action). In addition 
to the data points on which these scale function estimates are based, 
we include three data points calculated by L\"uscher \cite{L10} with 
the gradient method and two recent large lattice estimates of $\beta_t$ 
\cite{Fr13}. Summary and conclusions follow in the final 
section~\ref{sec_sum}.

\section{Definition and comparison of the scales} \label{sec_main}

Sommer \cite{So94} proposed to set a hadronic scale $r_i/a$ through the
force $F(r)$ between static quarks at intermediate distances $r$ by $r^2_i
F(r_i)=c_i$ (Sommer scale). For their SU(3) investigations NS \cite{NS02}
use the values
\begin{eqnarray} \label{r0r1}
  r^2_0\,F(r_0)\ =\ 1.65~~{\rm and}~~r^2_c\,F(r_c)\ =\ 0.65\,.
\end{eqnarray}
The $r_0$ value was suggested in the original paper by Sommer. It is 
used by NS for their smaller lattices and also by EHK, who employ 
also larger values for $c_i$, 
which we do not discuss here. 
The $r_c$ definition is used by NS for their set of large lattices. 
While a number of choices have to be made when calculating $r_i/a$ 
(for details see the EHK and NS papers), estimations of the 
deconfining transition temperatures $T_t=1/[a(\beta_t)N_t]$ are 
in essence free of ambiguities when one uses maxima of the Polyakov 
loop susceptibility on $N^3\,N_t$ lattices to determine $\beta_t(N_t)$ 
for the limit $N\to\infty$. In particular, when refining the lattice a 
switch of a reference value, like from $r_0$ to $r_c$ (\ref{r0r1}), is 
unwarranted when $T_t$ is used.

In the following we compile the analytical expressions of the three 
scaling functions. The EHK scaling function, the second of Eqs.~(4.4) 
in their paper \cite{EHK98} with $\hat{a}$ defined by their Eq.~(4.1), 
is given by
\begin{eqnarray} \label{EHK}
  \left[a\,\Lambda_L\right]^{EHK} = f^{EHK}_{\lambda}(\beta) = 
  \lambda^{EHK}(g^2)\,f^{as}_{\lambda}(g^2)\,,
\end{eqnarray}
and derived from data in the range $5.6\le\beta\le 6.5$. Here $f^{as}
(g^2)$ is the universal two-loop scaling function of SU(3) gauge theory,
\begin{eqnarray} \label{fas}
  f^{as}(g^2)\ =\ \left(b_0\,g^2\right)^{-b_1/(2b_0^2)}\,
  e^{-1/(2b_0\,g^2)}~~~{\rm with}~~~
  b_0\ =\ \frac{11}{3}\frac{3}{16\pi^2}\,,~~
  b_1\ =\ \frac{34}{3}\left(\frac{3}{16\pi^2}\right)^2\,.
\end{eqnarray}
Higher perturbative and non-perturbative corrections are 
parametrized by
\begin{eqnarray}
  \lambda^{EHK}(g^2)\ =\ (1+a_1\,\hat{a}^2+a_2\,\hat{a}^4)/a_0~~
  {\rm with}~~\hat{a}=\hat{a}(g^2) = f^{as}(g^2)/f^{as}(1) 
\end{eqnarray}
and the coefficients are given by $a_0=0.01596$, $a_1=0.2106$, $a_2
=0.05492$. Up to the over-all constant $1/a_0$, the asymptotic scale 
$f^{as}(g^2)$ is approached for $\beta\to\infty$. In contrast to 
that NS present their scale in form of a polynomial fit, Eq.~(2.6) 
in their paper \cite{NS02}, which is supposed to be valid in the 
region $5.7\le\beta\le 6.92$: $\left[a\,\Lambda_L\right]^{NS}=
f_{\lambda}^{NS}(\beta)$ with
\begin{eqnarray} \label{NS}
  f_{\lambda}^{NS}(\beta) = \exp\left[\,-1.6804-1.7331\,(\beta-6)
  + 0.7849\, (\beta-6)^2 - 0.4428\,(\beta-6)^3\ \right]\,.
\end{eqnarray}
The BBV scaling function, Eq.~(19) in their paper \cite{BBV06},  
is given by\footnote{To get convenient constants in the upcoming 
table~\ref{tab:c}, our definition (\ref{BBV}) differs by a factor 
10 from the one in \cite{BBV06}.}
\begin{eqnarray} \label{BBV}
  \left[a\,\Lambda_L\right]^{BBV} = f^{BBV}_{\lambda}(\beta) = 
  10\times\lambda^{BBV}(g^2)\,f^{as}(g^2)\,,
\end{eqnarray}
where $f^{as}$ is again the asymptotic scaling function (\ref{fas})
and higher perturbative and non-perturbative corrections are
parametrized by
\begin{eqnarray} 
  \lambda^{BBV}(g^2)\ =\ 1+e^{\ln a_1}\,e^{-a_2/g^2}+a_3\,g^2+a_4\,g^4
\end{eqnarray}
with the coefficients $\ln a_1=18.08596$, $a_2=19.48099$, $a_3=
-0.03772473$, $a_4=0.5089052.$ As the EHK scale, the BBV scale 
approaches up to a constant factor $f^{as}(g^2)$ for $\beta\to\infty$. 

\begin{table} \begin{center} \scalebox{1.0}{%
\begin{tabular}{|c|c|c|c|c|c|c|c|} \hline
$\beta$&EHK $r_0/a$&$\beta$&NS $r_0/a$&$\beta$&NS $r_c/a$&$\beta_t$&
Bielefeld $(aT_t)^{-1}$                                       \\ \hline 
5.60&2.344 (08) &5.70&2.922 (09)&6.57&6.25 (4)&5.6925 (05)$^*$&
4.0000 (18)                                                   \\ \hline 
5.70&2.990 (24) &5.80&3.673 (05)&6.69&7.29 (5)&5.8941 (05)&6.0000 (55)     
                                                              \\ \hline 
5.85&4.103 (12) &5.95&4.898 (12)&6.81&8.49 (5)&6.0609 (09)&8.000~~(12)     
                                                              \\ \hline 
6.00&5.3681 (86)&6.07&6.033 (17)&6.92&9.82 (6)&6.3331 (13)&12.000~(22)     
                                                              \\ \hline 
6.20&7.368 (30) &6.20&7.380 (26)& 
\multicolumn{2}{c|}{L\"uscher $\sqrt{8t_0}/a$}   &$-$ &$-$    \\ \hline 
6.20&7.368 (30) &6.20&7.380 (26)&5.96&4.7205 (53)& 
\multicolumn{2}{c|}{Francis et al. $(aT_t)^{-1}$}             \\ \hline
6.40&9.82~~(12) &6.40&9.74~~(05)&6.17&6.6266 (85)&6.4488 (59)&
14.00~~(12)                                                   \\ \hline 
6.50&11.23 (21) &$-$ &$-$       &6.42&9.4830 (97)&6.5509 (39)&
16.000~(82)                                                   \\ \hline 
\end{tabular}} \caption{Data used. $^*$The statistical error bar of 
this data point has been increased, so that it does not dominate the 
whole $T_c$ set, when the overall constant is adjusted to fit to the 
NS or EHK scale function. \label{tab:data}}
\end{center} \end{table}

In table~\ref{tab:data} data are compiled on which the scales rely. As 
usual error bars are given in parenthesis and apply to the last digits. 
The EHK data are from table~4 of their paper \cite{EHK98}, which 
includes also results from other groups. Thus several data point exists 
at some $\beta$, which are here combined into one estimate per $\beta$ 
value. Their $\beta=5.54$ data point is omitted, because it is not used 
for the determination of their $r_0/a$ scaling function (\ref{EHK}). The
NS data are from table~1 of their paper \cite{NS02}. The Bielefeld data 
are from table~2 of their paper \cite{Bo96}. We also list the three
gradient flow data points from L\"uscher \cite{L10} and two recent 
large-lattice $\beta_t$ estimates from Francis et al.\ \cite{Fr13}.
As these data are not used for the determination of the scaling functions 
they provide independent tests. The statistical errors for estimates of 
deconfining transition transition temperatures are in $\beta_t$ with 
$N_t$ fixed. To allow for direct comparison with the statistical 
accuracy of the Sommer method, we attach to $(aT_t)^{-1}$ error bars 
by means of the equation
\begin{eqnarray} \label{dTc}
  \triangle (aT_t)^{-1}\ =\ \frac{N_t}{f^{BBV}_{\lambda}(\beta_t)}\,
  \left[f^{BBV}_{\lambda}(\beta_t) - 
  f^{BBV}_{\lambda}(\beta_t-\triangle\beta_t)\right]\,.
\end{eqnarray}
%
\begin{table} \begin{center} \scalebox{1.0}{%
\begin{tabular}{|c|c|c|c|c|c|c|} \hline
 &    EHK $r_0$&  EHK $r_0-1$&      NS $r_0$&     NS $r_c$& 
     Bielefeld$+$& L\"uscher \\ \hline
E& 0.9994  (14)& 0.9996 (15) & 0.99204 (97) & 0.5172 (17) & 
     1.35102 (81)& 0.94272 (62) \\ \hline 
N& 1.0055  (14)& 1.0031 (15) & 0.99995 (98) & 0.5140 (17) & 
     1.36108 (81)& 0.94420 (62) \\ \hline 
B& 0.21566 (28)& 0.21646 (31)& 0.21415 (21) & 0.11024 (35)& 
     0.29146 (17)& 0.20388 (13) \\ \hline 
\end{tabular}} 
\caption{Scale constants $c$ from fitting Eq.~(2.10) to the data 
(E for EHK, N for NS and B for BBV).  
\label{tab:c}}
\end{center} \end{table}
\begin{table} \begin{center} \scalebox{1.0}{%
\begin{tabular}{|c|c|c|c|c|c|c|} \hline 
  & EHK $r_0$&EHK $r_0-1$&NS $r_0$&NS $r_c$&Bielefeld$+$ & L\"uscher
\\ \hline
EHK~~~~~~~~~~~~~& 0.83    &0.66     &$10^{-7}$&0.035&$10^{-15}$&
$10^{-3}$ \\ \hline 
NS~~~~~~~~~~~~~~&$10^{-6}$&$10^{-3}$& 0.12    &0.52 & 0        &
$10^{-8}$ \\ \hline 
BBV~~~~~~~~~~~~~&$10^{-9}$&0.45     & 0       &0.54 &0.31      & 
0.80      \\ \hline \end{tabular}} 
\caption{Probabilities $Q$ that the discrepancy between scale and data 
set is due to chance. Zero indicates a positive number smaller than
$10^{-12}$. \label{tab:Q}}
\end{center} \end{table}
For each of the the three scaling functions we perform one-parameter fits 
of the form
\begin{eqnarray} \label{cof}
  c / f_{\lambda} (\beta)
\end{eqnarray}
to altogether six data sets: EKH $r_0$ data, EHK $r_0$ data with the
data point for $\beta=5.6$ removed (the lowest $\beta$ entering the
determination of their scaling function) and denoted EHK $r_0-1$, NS 
$r_0$ data, NS $r_c$ data, combined Bielefeld and Francis et al.\ 
data denoted Bielefeld$+$ and L\"uscher's data points. The NS data 
are split, because their $r_0$ and $r_1$ data require independent 
determinations of the over all constant in (\ref{cof}), while the 
Bielefeld and Francis et al.\ data are combined by the opposite reason. 
The results for the twelve constants are compiled in table~\ref{tab:c}. 

Even more interesting than the constants are the thus obtained 
goodness of fit values $Q$, which are given in 
table~\ref{tab:Q}. We see that the EHK $r_0$ data are only consistent 
with the EHK scale, similarly the NS $r_0$ data are only consistent 
with the NS scale and the Bielefeld$+$ data only with the BBV scale. 
The NS $r_c$ data from large lattices are rather inaccurate. They are 
consistent with the NS and BBV scales and almost consistent with 
the EHK scale. Leaving the $\beta=5.6$ EHK data point out, because 
we may not expect universal scaling at such a small $\beta$ value, 
the EHK $r_0-1$ data are then consistent with the BBV scale, but still 
in disagreement with the NS scale. In the last column it is seen that 
only the BBV scale is consistent with L\"uscher's data points.

\begin{figure} \begin{center} 
\includegraphics[width=0.74\textwidth]{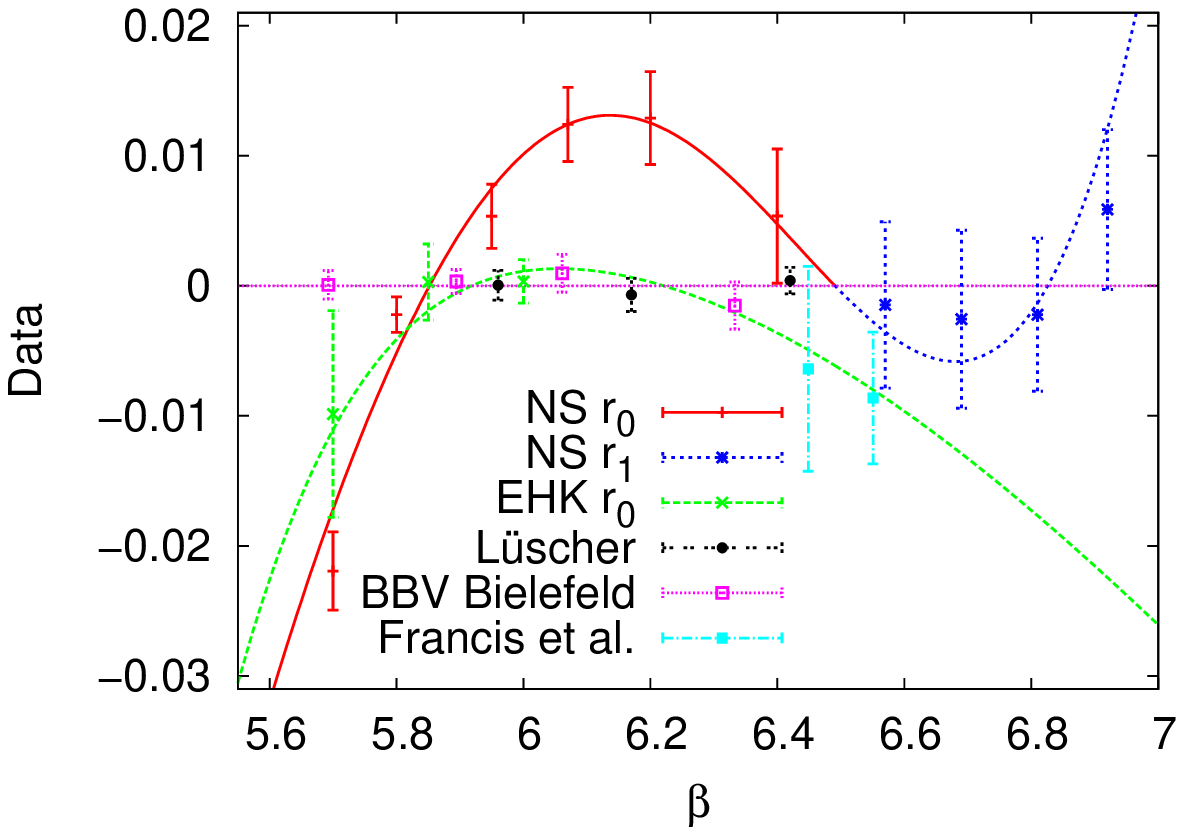} 
\caption{Relative deviations after the best fit of each 
data set to the BBV scale function.} \label{fig_bbv}
\includegraphics[width=0.74\textwidth]{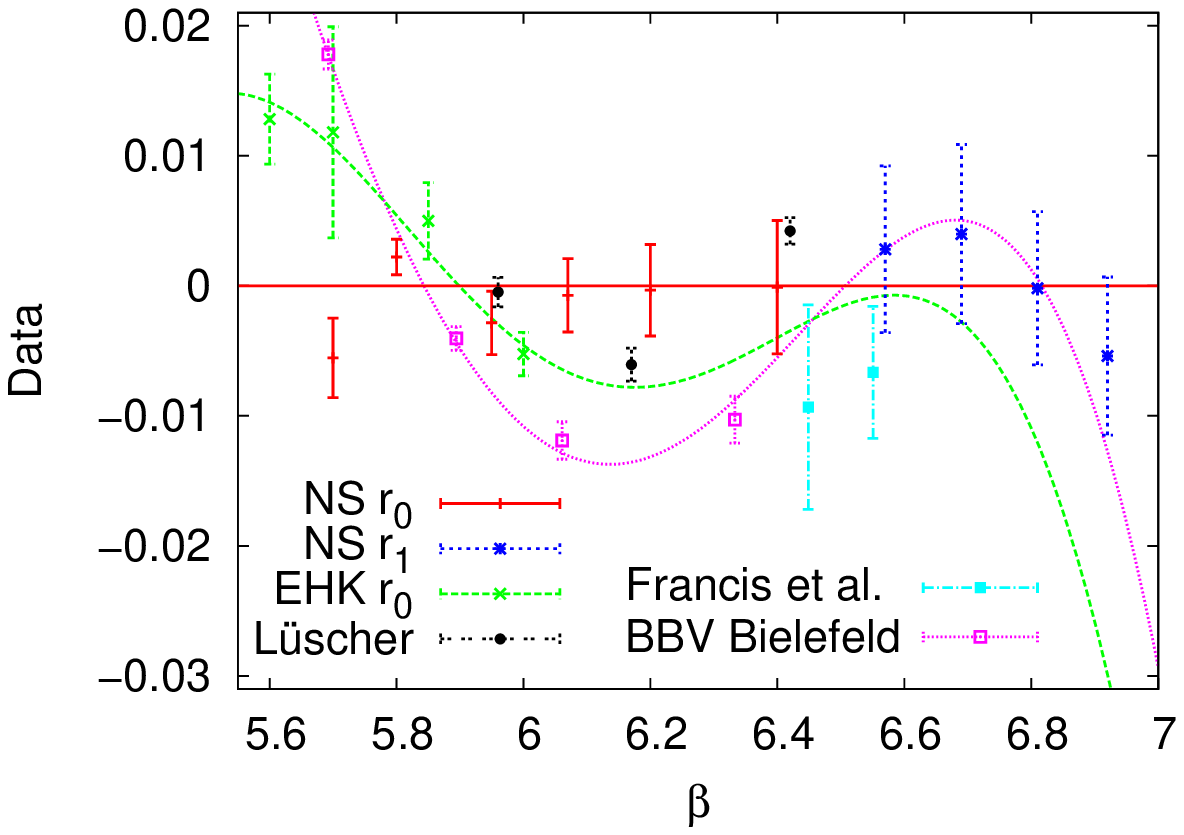} 
\caption{Relative deviations after the best fit of each 
data set to the NS scale.} \label{fig_ns}
\end{center} \end{figure} 

\begin{figure} \begin{center} 
\includegraphics[width=0.74\textwidth]{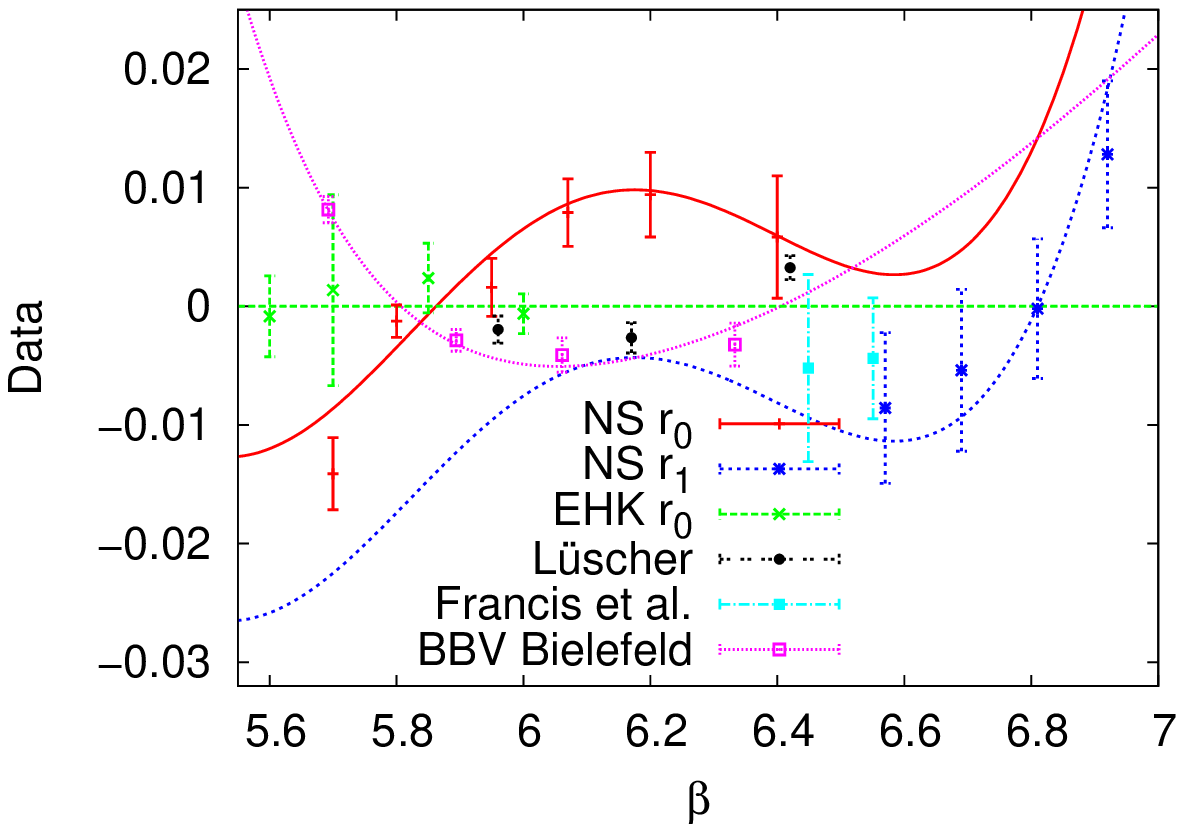} 
\caption{Relative deviations after the best fit of each 
data set to the EHK scale function.} \label{fig_ehk}
\end{center} \end{figure} 

Using the best fits to the BBV scale, regardless of good or bad $Q$
values, Fig.~\ref{fig_bbv} is obtained for the differences between 
the data and the BBV scale function divided by this function (relative 
deviation). Correspondingly, the relative deviations to the EHK and NS 
scale functions are calculated and shown in the figure. Rotating the 
scale functions around, the relative deviations from the NS and EHK 
scales are found in the same way and shown in Figs.~\ref{fig_ns} 
and~\ref{fig_ehk}.

The ratio between the NS data sets $r_0$ and $r_c$ changes when 
different scale functions are used. From the constants of 
table~\ref{tab:c} one finds
\begin{eqnarray} \label{rBBV}
  (r_c/r_0)^{BBV} &=& 0.11024~(35)/0.21415~(21)\ =\  0.5148~(18)\,,
  \\ \label{rNS}
   (r_c/r_0)^{NS} &=& 0.5140~(17)/0.99995~(98)\ =\ 0.5140~(18)\,,
  \\ \label{rEHK}
  (r_c/r_0)^{EHK} &=& 0.5172~(17)/0.99204~(97)\ =\ 0.5214~(18)\,.
\end{eqnarray}
The first values for $(r_c/r_0)^{BBV}$ and $(r_c/r_0)^{NS}$ are in 
statistical agreement with one another as well as with the ratio 
$r_c/r_0=0.5133~(24)$, which is given in Eq.~(2.5) of the NS paper 
and used to determine the NS scale function. For $(r_c/r_0)^{NS}$ this 
is obvious in Fig.~\ref{fig_ns}, where the NS scale function (i.e., 
the zero-line) fits both NS data sets well. All other data are 
in disagreement with this scale. The BBV reference scale of 
Fig.~\ref{fig_bbv} fits the EHK $r_0-1$ data\footnote{The fit drawn 
is for the EHK $r_0$ data. It becomes good for the EHK $r_0-1$ data 
(omission of the $\beta=5.6$ data point implies also small changes 
for the EHK data coefficients as listed in table~\ref{tab:c}).}, 
the $T_c$, the NS $r_c$ and L\"uscher's
data well and is in disagreement with the NS $r_0$ data and the $\beta
=5.6$ EHK value. Due to the slight difference between the ratios 
(\ref{rBBV}) and (\ref{rNS}) the NS scale on its $r_0$ data should 
in Fig.~\ref{fig_bbv} be slightly higher than the NS scale on its 
$r_c$ data. As this stays within statistical errors, we have just 
averaged the two curves, but use distinct colors, red for the $r_0$ 
and blue for the $r_c$ range. Such averaging is not possible when 
plotting the NS data versus the EHK scale, because the ratio 
(\ref{rEHK}) is incompatible with the other two ratios. It amounts 
to the difference between the red and blue curves in Fig.~\ref{fig_ehk}.

\begin{figure} \begin{center} 
\includegraphics[width=0.74\textwidth]{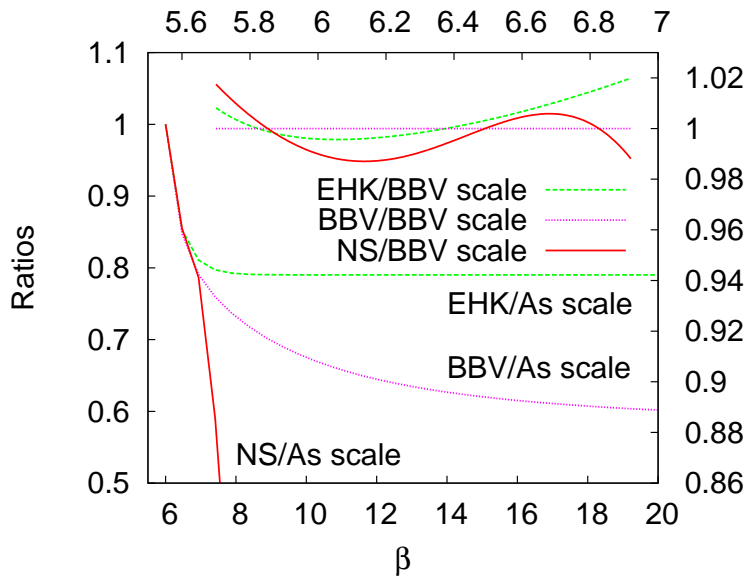} 
\caption{Ratios with respect to the BBV scale (upper part, top abscissa 
and right ordinate) and asymptotic behavior of the scales (lower part, 
bottom abscissa and left ordinate).} \label{fig_asscales}
\end{center} \end{figure} 

\section{Summary and conclusions} \label{sec_sum}

Table~\ref{tab:Q} shows that the three scale functions (EHK, NS and BBV) 
are derived from data sets, given in table~\ref{tab:data}, which are 
mutually inconsistent in the range up to $\beta=6.4$, while the NS $r_c$ 
data for the range $6.57 \le\beta\le 6.92$ are not very restrictive. 
Only the BBV scaling function is consistent with L\"uscher's accurate 
data (see also Figs.~\ref{fig_bbv} to~\ref{fig_ehk}).

In the range $5.65\le\beta\le 6.92$ the relative discrepancy between 
the scales is never larger than $\pm 2$\% as is shown in the upper part 
of Fig.~\ref{fig_asscales} for ratios of the form $\,const\,
f^{EHK}_{\lambda}/f^{BBV}_{\lambda}$ and $\,const'\,f^{NS}_{\lambda}/
f^{BBV}_{\lambda}$ (the upper abscissa and the right ordinate apply and 
the constants (\ref{cof}) used from table~\ref{tab:c} are the same as 
those for Fig.~\ref{fig_bbv}). Note that the previous figures, which 
exhibit relative deviations from the scales, cover a corresponding 
[-0.03:0.02] range. 

The lower part of Fig.~\ref{fig_asscales} shows that the EHK and BBV 
scales approach the universal asymptotic scale (\ref{fas}) in rather 
distinct ways, whereas such a parametrization is not attempted by NS 
(this part of the figure uses a normalization in which all scales agree 
at $\beta =6$). The discrepancy between EHK and BBV with respect to
the approach of the asymptotic scale relies on making distinct assumptions
which can only be resolved on the basis of more accurate results at 
larger $\beta$ values, which could come from calculations of the SU(3) 
deconfining temperature for $N_t>12$. This may need some innovative 
techniques as the $N_t=14$ and 16 data from Francis et al.\ are seen 
to exhibit similar inaccuracies as the large lattice NS data. Most 
promising may be calculations with the gradient method at larger 
$\beta$ values. That this will work is also not obvious. For instance, 
the sensitivity of the gradient method to topological excitations 
\cite{L10,BE14} on periodic lattices turns into a disadvantage when 
it comes to accurate scale calculations. 
\medskip

\noindent
{\bf Acknowledgments:} This work has in part been supported by DOE grant 
DE-FG02-97ER-41022 and by NERSC ERCAP 86977 and ERCAP 86979. I like to 
thank Urs Heller for useful comments on the manuscript.

\end{document}